\pdfoutput=1

\documentclass[preprint
  ,prd
  ,tightenlines
  ,superscriptaddress
]{revtex4-1}

\usepackage{graphicx}
\usepackage{dcolumn}
\usepackage{color}
\usepackage{multirow}
\usepackage{placeins}
\usepackage{caption}
\usepackage{amsmath}
\usepackage{amssymb}
\usepackage{amsfonts}
\usepackage{upgreek}
\usepackage{xspace}
\usepackage{url}
\usepackage{hyperref}

\input{belle2sym.tex}

\begin{document}

\vspace*{0.5\baselineskip}
\resizebox{!}{3cm}{\includegraphics{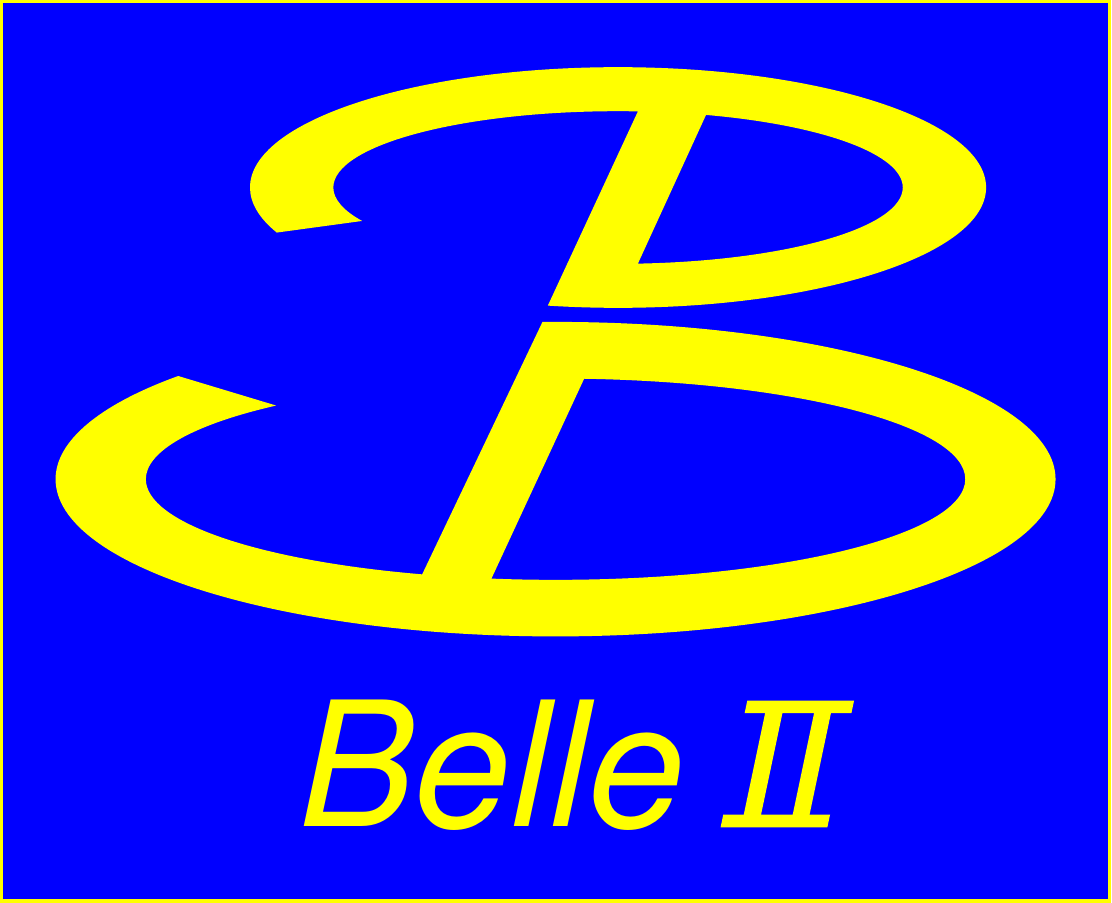}}

\vspace*{-4\baselineskip}
\begin{flushright}
BELLE2-CONF-PH-2020-003\\
\end{flushright}
\vspace*{2.5\baselineskip}

\title{Measurement of the \Bd lifetime using fully reconstructed hadronic decays in the 2019 Belle~II dataset}

\newcommand{\instSinica}{Academia Sinica, Taipei 11529, Taiwan}
\newcommand{\instBeihang}{Beihang University, Beijing 100191, China}
\newcommand{\instBUAP}{Benemerita Universidad Autonoma de Puebla, Puebla 72570, Mexico}
\newcommand{\instBNL}{Brookhaven National Laboratory, Upton, New York 11973, U.S.A.}
\newcommand{\instBINP}{Budker Institute of Nuclear Physics SB RAS, Novosibirsk 630090, Russian Federation}
\newcommand{\instCMU}{Carnegie Mellon University, Pittsburgh, Pennsylvania 15213, U.S.A.}
\newcommand{\instCPPM}{Centre de Physique des Particules de Marseille, 13288 Marseille, France}
\newcommand{\instCinvestavIPN}{Centro de Investigacion y de Estudios Avanzados del Instituto Politecnico Nacional, Mexico City 07360, Mexico}
\newcommand{\instPrague}{Faculty of Mathematics and Physics, Charles University, 121 16 Prague, Czech Republic}
\newcommand{\instChiangMai}{Chiang Mai University, Chiang Mai 50202, Thailand}
\newcommand{\instChiba}{Chiba University, Chiba 263-8522, Japan}
\newcommand{\instChonnam}{Chonnam National University, Gwangju 61186, South Korea}
\newcommand{\instConacyt}{Consejo Nacional de Ciencia y Tecnolog\'{\i}a, Mexico City 03940, Mexico}
\newcommand{\instDESY}{Deutsches Elektronen--Synchrotron, 22607 Hamburg, Germany}
\newcommand{\instDuke}{Duke University, Durham, North Carolina 27708, U.S.A.}
\newcommand{\instDuyTan}{Institute of Theoretical and Applied Research (ITAR), Duy Tan University, Hanoi 100000, Vietnam}
\newcommand{\instENEA}{ENEA Casaccia, I-00123 Roma, Italy}
\newcommand{\instEri}{Earthquake Research Institute, University of Tokyo, Tokyo 113-0032, Japan}
\newcommand{\instJuelich}{Forschungszentrum J\"{u}lich, 52425 J\"{u}lich, Germany}
\newcommand{\instFuJen}{Department of Physics, Fu Jen Catholic University, Taipei 24205, Taiwan}
\newcommand{\instFudan}{Key Laboratory of Nuclear Physics and Ion-beam Application (MOE) and Institute of Modern Physics, Fudan University, Shanghai 200443, China}
\newcommand{\instGoettingen}{II. Physikalisches Institut, Georg-August-Universit\"{a}t G\"{o}ttingen, 37073 G\"{o}ttingen, Germany}
\newcommand{\instGifu}{Gifu University, Gifu 501-1193, Japan}
\newcommand{\instSOKENDAI}{The Graduate University for Advanced Studies (SOKENDAI), Hayama 240-0193, Japan}
\newcommand{\instGyeongsang}{Gyeongsang National University, Jinju 52828, South Korea}
\newcommand{\instHanyang}{Department of Physics and Institute of Natural Sciences, Hanyang University, Seoul 04763, South Korea}
\newcommand{\instKEK}{High Energy Accelerator Research Organization (KEK), Tsukuba 305-0801, Japan}
\newcommand{\instJPARC}{J-PARC Branch, KEK Theory Center, High Energy Accelerator Research Organization (KEK), Tsukuba 305-0801, Japan}
\newcommand{\instHSE}{Higher School of Economics (HSE), Moscow 101000, Russian Federation}
\newcommand{\instIISER}{Indian Institute of Science Education and Research Mohali, SAS Nagar, 140306, India}
\newcommand{\instIITBhubaneswar}{Indian Institute of Technology Bhubaneswar, Satya Nagar 751007, India}
\newcommand{\instIITGuwahati}{Indian Institute of Technology Guwahati, Assam 781039, India}
\newcommand{\instIITHyderabad}{Indian Institute of Technology Hyderabad, Telangana 502285, India}
\newcommand{\instIITMadras}{Indian Institute of Technology Madras, Chennai 600036, India}
\newcommand{\instIndiana}{Indiana University, Bloomington, Indiana 47408, U.S.A.}
\newcommand{\instIHEPRussia}{Institute for High Energy Physics, Protvino 142281, Russian Federation}
\newcommand{\instHEPHYVienna}{Institute of High Energy Physics, Vienna 1050, Austria}
\newcommand{\instIHEPChina}{Institute of High Energy Physics, Chinese Academy of Sciences, Beijing 100049, China}
\newcommand{\instChennai}{Institute of Mathematical Sciences, Chennai 600113, India}
\newcommand{\instIPP}{Institute of Particle Physics (Canada), Victoria, British Columbia V8W 2Y2, Canada}
\newcommand{\instIOP}{Institute of Physics, Vietnam Academy of Science and Technology (VAST), Hanoi, Vietnam}
\newcommand{\instIFIC}{Instituto de Fisica Corpuscular, Paterna 46980, Spain}
\newcommand{\instFrascati}{INFN Laboratori Nazionali di Frascati, I-00044 Frascati, Italy}
\newcommand{\instNapoliINFN}{INFN Sezione di Napoli, I-80126 Napoli, Italy}
\newcommand{\instPadovaINFN}{INFN Sezione di Padova, I-35131 Padova, Italy}
\newcommand{\instPerugiaINFN}{INFN Sezione di Perugia, I-06123 Perugia, Italy}
\newcommand{\instPisaINFN}{INFN Sezione di Pisa, I-56127 Pisa, Italy}
\newcommand{\instRomaINFN}{INFN Sezione di Roma, I-00185 Roma, Italy}
\newcommand{\instRomaTreINFN}{INFN Sezione di Roma Tre, I-00146 Roma, Italy}
\newcommand{\instTorinoINFN}{INFN Sezione di Torino, I-10125 Torino, Italy}
\newcommand{\instTriesteINFN}{INFN Sezione di Trieste, I-34127 Trieste, Italy}
\newcommand{\instJAEA}{Advanced Science Research Center, Japan Atomic Energy Agency, Naka 319-1195, Japan}
\newcommand{\instMainz}{Johannes Gutenberg-Universit\"{a}t Mainz, Institut f\"{u}r Kernphysik, D-55099 Mainz, Germany}
\newcommand{\instGiessen}{Justus-Liebig-Universit\"{a}t Gie\ss{}en, 35392 Gie\ss{}en, Germany}
\newcommand{\instKarlsruhe}{Institut f\"{u}r Experimentelle Teilchenphysik, Karlsruher Institut f\"{u}r Technologie, 76131 Karlsruhe, Germany}
\newcommand{\instKennesaw}{Kennesaw State University, Kennesaw, Georgia 30144, U.S.A.}
\newcommand{\instKitasato}{Kitasato University, Sagamihara 252-0373, Japan}
\newcommand{\instKISTI}{Korea Institute of Science and Technology Information, Daejeon 34141, South Korea}
\newcommand{\instKorea}{Korea University, Seoul 02841, South Korea}
\newcommand{\instKSU}{Kyoto Sangyo University, Kyoto 603-8555, Japan}
\newcommand{\instKyotoU}{Kyoto University, Kyoto 606-8501, Japan}
\newcommand{\instKyungpook}{Kyungpook National University, Daegu 41566, South Korea}
\newcommand{\instLAL}{Universit\'{e} Paris-Saclay, CNRS/IN2P3, IJCLab, 91405 Orsay, France}
\newcommand{\instLPI}{P.N. Lebedev Physical Institute of the Russian Academy of Sciences, Moscow 119991, Russian Federation}
\newcommand{\instLNNU}{Liaoning Normal University, Dalian 116029, China}
\newcommand{\instLMU}{Ludwig Maximilians University, 80539 Munich, Germany}
\newcommand{\instLuther}{Luther College, Decorah, Iowa 52101, U.S.A.}
\newcommand{\instMNITJaipur}{Malaviya National Institute of Technology Jaipur, Jaipur 302017, India}
\newcommand{\instMPP}{Max-Planck-Institut f\"{u}r Physik, 80805 M\"{u}nchen, Germany}
\newcommand{\instMPGHLL}{Semiconductor Laboratory of the Max Planck Society, 81739 M\"{u}nchen, Germany}
\newcommand{\instMcGill}{McGill University, Montr\'{e}al, Qu\'{e}bec, H3A 2T8, Canada}
\newcommand{\instMETU}{Middle East Technical University, 06531 Ankara, Turkey}
\newcommand{\instMEPhI}{Moscow Physical Engineering Institute, Moscow 115409, Russian Federation}
\newcommand{\instNagoya}{Graduate School of Science, Nagoya University, Nagoya 464-8602, Japan}
\newcommand{\instNagoyaKMI}{Kobayashi-Maskawa Institute, Nagoya University, Nagoya 464-8602, Japan}
\newcommand{\instNaraWu}{Nara Women's University, Nara 630-8506, Japan}
\newcommand{\instUNAM}{National Autonomous University of Mexico, Mexico City, Mexico}
\newcommand{\instNTUTaiwan}{Department of Physics, National Taiwan University, Taipei 10617, Taiwan}
\newcommand{\instNUUTaiwan}{National United University, Miao Li 36003, Taiwan}
\newcommand{\instKrakow}{H. Niewodniczanski Institute of Nuclear Physics, Krakow 31-342, Poland}
\newcommand{\instNiigata}{Niigata University, Niigata 950-2181, Japan}
\newcommand{\instNSU}{Novosibirsk State University, Novosibirsk 630090, Russian Federation}
\newcommand{\instOkinawa}{Okinawa Institute of Science and Technology, Okinawa 904-0495, Japan}
\newcommand{\instOsakaCity}{Osaka City University, Osaka 558-8585, Japan}
\newcommand{\instRCNP}{Research Center for Nuclear Physics, Osaka University, Osaka 567-0047, Japan}
\newcommand{\instPNNL}{Pacific Northwest National Laboratory, Richland, Washington 99352, U.S.A.}
\newcommand{\instPanjab}{Panjab University, Chandigarh 160014, India}
\newcommand{\instPeking}{Peking University, Beijing 100871, China}
\newcommand{\instPanjabPAU}{Punjab Agricultural University, Ludhiana 141004, India}
\newcommand{\instRIKEN}{Theoretical Research Division, Nishina Center, RIKEN, Saitama 351-0198, Japan}
\newcommand{\instXavier}{St. Francis Xavier University, Antigonish, Nova Scotia, B2G 2W5, Canada}
\newcommand{\instSeoul}{Seoul National University, Seoul 08826, South Korea}
\newcommand{\instShandong}{Shandong University, Jinan 250100, China}
\newcommand{\instSPU}{Showa Pharmaceutical University, Tokyo 194-8543, Japan}
\newcommand{\instSoochow}{Soochow University, Suzhou 215006, China}
\newcommand{\instSoongsil}{Soongsil University, Seoul 06978, South Korea}
\newcommand{\instLjubljanaJSI}{J. Stefan Institute, 1000 Ljubljana, Slovenia}
\newcommand{\instKyiv}{Taras Shevchenko National Univ. of Kiev, Kiev, Ukraine}
\newcommand{\instTata}{Tata Institute of Fundamental Research, Mumbai 400005, India}
\newcommand{\instTUM}{Department of Physics, Technische Universit\"{a}t M\"{u}nchen, 85748 Garching, Germany}
\newcommand{\instECUTUM}{Excellence Cluster Universe, Technische Universit\"{a}t M\"{u}nchen, 85748 Garching, Germany}
\newcommand{\instTelAviv}{Tel Aviv University, School of Physics and Astronomy, Tel Aviv, 69978, Israel}
\newcommand{\instToho}{Toho University, Funabashi 274-8510, Japan}
\newcommand{\instTohoku}{Department of Physics, Tohoku University, Sendai 980-8578, Japan}
\newcommand{\instTitech}{Tokyo Institute of Technology, Tokyo 152-8550, Japan}
\newcommand{\instTokyoMetropolitan}{Tokyo Metropolitan University, Tokyo 192-0397, Japan}
\newcommand{\instUAS}{Universidad Autonoma de Sinaloa, Sinaloa 80000, Mexico}
\newcommand{\instNapoliUNIV}{Dipartimento di Scienze Fisiche, Universit\`{a} di Napoli Federico II, I-80126 Napoli, Italy}
\newcommand{\instNapoliUNIVA}{Dipartimento di Agraria, Universit\`{a} di Napoli Federico II, I-80055 Portici (NA), Italy}
\newcommand{\instPadovaUNIV}{Dipartimento di Fisica e Astronomia, Universit\`{a} di Padova, I-35131 Padova, Italy}
\newcommand{\instPerugiaUNIV}{Dipartimento di Fisica, Universit\`{a} di Perugia, I-06123 Perugia, Italy}
\newcommand{\instPisaUNIV}{Dipartimento di Fisica, Universit\`{a} di Pisa, I-56127 Pisa, Italy}
\newcommand{\instRomaUNIV}{Universit\`{a} di Roma ``La Sapienza,'' I-00185 Roma, Italy}
\newcommand{\instRomaTreUNIV}{Dipartimento di Matematica e Fisica, Universit\`{a} di Roma Tre, I-00146 Roma, Italy}
\newcommand{\instTorinoUNIV}{Dipartimento di Fisica, Universit\`{a} di Torino, I-10125 Torino, Italy}
\newcommand{\instTriesteUNIV}{Dipartimento di Fisica, Universit\`{a} di Trieste, I-34127 Trieste, Italy}
\newcommand{\instMontreal}{Universit\'{e} de Montr\'{e}al, Physique des Particules, Montr\'{e}al, Qu\'{e}bec, H3C 3J7, Canada}
\newcommand{\instIPHC}{Universit\'{e} de Strasbourg, CNRS, IPHC, UMR 7178, 67037 Strasbourg, France}
\newcommand{\instAdelaide}{Department of Physics, University of Adelaide, Adelaide, South Australia 5005, Australia}
\newcommand{\instBonn}{University of Bonn, 53115 Bonn, Germany}
\newcommand{\instUBC}{University of British Columbia, Vancouver, British Columbia, V6T 1Z1, Canada}
\newcommand{\instCincinnati}{University of Cincinnati, Cincinnati, Ohio 45221, U.S.A.}
\newcommand{\instFlorida}{University of Florida, Gainesville, Florida 32611, U.S.A.}
\newcommand{\instHamburg}{University of Hamburg, 20148 Hamburg, Germany}
\newcommand{\instHawaii}{University of Hawaii, Honolulu, Hawaii 96822, U.S.A.}
\newcommand{\instHeidelberg}{University of Heidelberg, 68131 Mannheim, Germany}
\newcommand{\instLjubljanaUniLJ}{Faculty of Mathematics and Physics, University of Ljubljana, 1000 Ljubljana, Slovenia}
\newcommand{\instLouisville}{University of Louisville, Louisville, Kentucky 40292, U.S.A.}
\newcommand{\instMalaya}{National Centre for Particle Physics, University Malaya, 50603 Kuala Lumpur, Malaysia}
\newcommand{\instLjubljanaUM}{University of Maribor, 2000 Maribor, Slovenia}
\newcommand{\instMelbourne}{School of Physics, University of Melbourne, Victoria 3010, Australia}
\newcommand{\instMississippi}{University of Mississippi, University, Mississippi 38677, U.S.A.}
\newcommand{\instUOM}{University of Miyazaki, Miyazaki 889-2192, Japan}
\newcommand{\instNovaGorica}{University of Nova Gorica, 5000 Nova Gorica, Slovenia}
\newcommand{\instPittsburgh}{University of Pittsburgh, Pittsburgh, Pennsylvania 15260, U.S.A.}
\newcommand{\instUSTC}{University of Science and Technology of China, Hefei 230026, China}
\newcommand{\instSAlabama}{University of South Alabama, Mobile, Alabama 36688, U.S.A.}
\newcommand{\instSCarolina}{University of South Carolina, Columbia, South Carolina 29208, U.S.A.}
\newcommand{\instSydney}{School of Physics, University of Sydney, New South Wales 2006, Australia}
\newcommand{\instTabuk}{Department of Physics, Faculty of Science, University of Tabuk, Tabuk 71451, Saudi Arabia}
\newcommand{\instUTokyo}{Department of Physics, University of Tokyo, Tokyo 113-0033, Japan}
\newcommand{\instIPMU}{Kavli Institute for the Physics and Mathematics of the Universe (WPI), University of Tokyo, Kashiwa 277-8583, Japan}
\newcommand{\instVictoria}{University of Victoria, Victoria, British Columbia, V8W 3P6, Canada}
\newcommand{\instVPI}{Virginia Polytechnic Institute and State University, Blacksburg, Virginia 24061, U.S.A.}
\newcommand{\instWayneState}{Wayne State University, Detroit, Michigan 48202, U.S.A.}
\newcommand{\instYamagata}{Yamagata University, Yamagata 990-8560, Japan}
\newcommand{\instYerevan}{Alikhanyan National Science Laboratory, Yerevan 0036, Armenia}
\newcommand{\instYonsei}{Yonsei University, Seoul 03722, South Korea}
\affiliation{\instSinica}
\affiliation{\instBeihang}
\affiliation{\instBUAP}
\affiliation{\instBNL}
\affiliation{\instBINP}
\affiliation{\instCMU}
\affiliation{\instCPPM}
\affiliation{\instCinvestavIPN}
\affiliation{\instPrague}
\affiliation{\instChiangMai}
\affiliation{\instChiba}
\affiliation{\instChonnam}
\affiliation{\instConacyt}
\affiliation{\instDESY}
\affiliation{\instDuke}
\affiliation{\instDuyTan}
\affiliation{\instENEA}
\affiliation{\instEri}
\affiliation{\instJuelich}
\affiliation{\instFuJen}
\affiliation{\instFudan}
\affiliation{\instGoettingen}
\affiliation{\instGifu}
\affiliation{\instSOKENDAI}
\affiliation{\instGyeongsang}
\affiliation{\instHanyang}
\affiliation{\instKEK}
\affiliation{\instJPARC}
\affiliation{\instHSE}
\affiliation{\instIISER}
\affiliation{\instIITBhubaneswar}
\affiliation{\instIITGuwahati}
\affiliation{\instIITHyderabad}
\affiliation{\instIITMadras}
\affiliation{\instIndiana}
\affiliation{\instIHEPRussia}
\affiliation{\instHEPHYVienna}
\affiliation{\instIHEPChina}
\affiliation{\instChennai}
\affiliation{\instIPP}
\affiliation{\instIOP}
\affiliation{\instIFIC}
\affiliation{\instFrascati}
\affiliation{\instNapoliINFN}
\affiliation{\instPadovaINFN}
\affiliation{\instPerugiaINFN}
\affiliation{\instPisaINFN}
\affiliation{\instRomaINFN}
\affiliation{\instRomaTreINFN}
\affiliation{\instTorinoINFN}
\affiliation{\instTriesteINFN}
\affiliation{\instJAEA}
\affiliation{\instMainz}
\affiliation{\instGiessen}
\affiliation{\instKarlsruhe}
\affiliation{\instKennesaw}
\affiliation{\instKitasato}
\affiliation{\instKISTI}
\affiliation{\instKorea}
\affiliation{\instKSU}
\affiliation{\instKyotoU}
\affiliation{\instKyungpook}
\affiliation{\instLAL}
\affiliation{\instLPI}
\affiliation{\instLNNU}
\affiliation{\instLMU}
\affiliation{\instLuther}
\affiliation{\instMNITJaipur}
\affiliation{\instMPP}
\affiliation{\instMPGHLL}
\affiliation{\instMcGill}
\affiliation{\instMETU}
\affiliation{\instMEPhI}
\affiliation{\instNagoya}
\affiliation{\instNagoyaKMI}
\affiliation{\instNaraWu}
\affiliation{\instUNAM}
\affiliation{\instNTUTaiwan}
\affiliation{\instNUUTaiwan}
\affiliation{\instKrakow}
\affiliation{\instNiigata}
\affiliation{\instNSU}
\affiliation{\instOkinawa}
\affiliation{\instOsakaCity}
\affiliation{\instRCNP}
\affiliation{\instPNNL}
\affiliation{\instPanjab}
\affiliation{\instPeking}
\affiliation{\instPanjabPAU}
\affiliation{\instRIKEN}
\affiliation{\instXavier}
\affiliation{\instSeoul}
\affiliation{\instShandong}
\affiliation{\instSPU}
\affiliation{\instSoochow}
\affiliation{\instSoongsil}
\affiliation{\instLjubljanaJSI}
\affiliation{\instKyiv}
\affiliation{\instTata}
\affiliation{\instTUM}
\affiliation{\instECUTUM}
\affiliation{\instTelAviv}
\affiliation{\instToho}
\affiliation{\instTohoku}
\affiliation{\instTitech}
\affiliation{\instTokyoMetropolitan}
\affiliation{\instUAS}
\affiliation{\instNapoliUNIV}
\affiliation{\instPadovaUNIV}
\affiliation{\instPerugiaUNIV}
\affiliation{\instPisaUNIV}
\affiliation{\instRomaUNIV}
\affiliation{\instRomaTreUNIV}
\affiliation{\instTorinoUNIV}
\affiliation{\instTriesteUNIV}
\affiliation{\instMontreal}
\affiliation{\instIPHC}
\affiliation{\instAdelaide}
\affiliation{\instBonn}
\affiliation{\instUBC}
\affiliation{\instCincinnati}
\affiliation{\instFlorida}
\affiliation{\instHamburg}
\affiliation{\instHawaii}
\affiliation{\instHeidelberg}
\affiliation{\instLjubljanaUniLJ}
\affiliation{\instLouisville}
\affiliation{\instMalaya}
\affiliation{\instLjubljanaUM}
\affiliation{\instMelbourne}
\affiliation{\instMississippi}
\affiliation{\instUOM}
\affiliation{\instNovaGorica}
\affiliation{\instPittsburgh}
\affiliation{\instUSTC}
\affiliation{\instSAlabama}
\affiliation{\instSCarolina}
\affiliation{\instSydney}
\affiliation{\instTabuk}
\affiliation{\instUTokyo}
\affiliation{\instIPMU}
\affiliation{\instVictoria}
\affiliation{\instVPI}
\affiliation{\instWayneState}
\affiliation{\instYamagata}
\affiliation{\instYerevan}
\affiliation{\instYonsei}
  \author{F.~Abudin{\'e}n}\affiliation{\instTriesteINFN} 
  \author{I.~Adachi}\affiliation{\instKEK}\affiliation{\instSOKENDAI} 
  \author{R.~Adak}\affiliation{\instFudan} 
  \author{K.~Adamczyk}\affiliation{\instKrakow} 
  \author{P.~Ahlburg}\affiliation{\instBonn} 
  \author{J.~K.~Ahn}\affiliation{\instKorea} 
  \author{H.~Aihara}\affiliation{\instUTokyo} 
  \author{N.~Akopov}\affiliation{\instYerevan} 
  \author{A.~Aloisio}\affiliation{\instNapoliUNIV}\affiliation{\instNapoliINFN} 
  \author{F.~Ameli}\affiliation{\instRomaINFN} 
% ...
\collaboration{Belle II Collaboration}

\author{The Belle II Collaboration}
\noaffiliation

\begin{abstract}
This document presents the measurement of \Bz meson lifetimes using  the 2019 Belle~II dataset that corresponds to an integrated luminosity of  $8.7 \pm 0.2$~\invfb. Each candidate is fully reconstructed with hadronic decay final states on the signal side, while the rest-of-event technique allows to infer the decay vertex position on the other (tag) side. \Bz lifetime is extracted from an unbinned maximum likelihood fit to the distribution of the difference between the signal side \Bz candidate and the tag side decay times. The measured lifetime is $\tau_{\Bd} = 1.48 \pm 0.28 \pm 0.06 $ ps, where the first uncertainty is statistical and the second is systematic.

\end{abstract}

\maketitle

{\renewcommand{\thefootnote}{\fnsymbol{footnote}}}
\setcounter{footnote}{0}

\section{Introduction}

This document summarizes the \Bd lifetime measurement from hadronic decays using the Belle~II experiment early dataset. The analysis demonstrates the use of vertex fitting tools and time resolution modeling. We describe the data sets used, the selection of signal candidates and the fit strategy to extract the lifetime parameter $\tau_{\Bd}$. Finally, we present the results along with an evaluation of the systematic uncertainties.

\section{Belle II detector and data sets used}
\label{sec:dataset}

The  Belle~II  detector~\cite{Abe:2010gxa,10.1093/ptep/ptz106}  operates  at  the  SuperKEKB  asymmetric-energy electron-positron  collider~\cite{Superkekb}, located  at  the  KEK  laboratory  in  Tsukuba,  Japan. The  detector consists of several subdetectors arranged around the beam pipe in a cylindrical geometry. The innermost subdetector is the vertex detector (VXD), which includes two layers of silicon pixel detectors~\cite{PXD} and four outer layers of silicon strip detectors~\cite{SVD}. Currently, the second pixel layer is installed in only a small part of the solid angle, and the remaining VXD layers are fully installed.  Most of the tracking volume is consists of a helium-based, small-cell central drift chamber (CDC). Outside the CDC, a Cherenkov-light angle and time-of-propagation detector (TOP) provides charged-particle identification in the barrel region. In the forward endcap, this function is provided by a proximity-focusing, aerogel ring imaging Cherenkov detector (ARICH). An electromagnetic  calorimeter  (ECL)  consists  of  a  barrel  and  two  endcap  sections  made  of CsI(Tl) crystals. A uniform 1.5 T magnetic field is provided by a solenoid situated outside the ECL. Multiple layers of scintillators and resistive plate chambers, inserted between the magnetic flux-return iron plates, constitute the $K_{L}$ and muon (KLM) identification system.

The  data  used  in  this  analysis  were  collected  at  a  center-of-mass  energy  of 10.58 GeV, corresponding to the mass of the \FourS resonance.  The energies of the electron and positron beams are 7 GeV and 4 GeV, respectively, resulting in a boost of $\beta \gamma = 0.28$ of the \FourS relative to the lab frame. The integrated luminosity used in this analysis is $8.7 \pm 0.2$~\invfb~\cite{Abudinen:2019osb}.

We use also a generic simulated Monte Carlo (MC) sample that includes \epem \to \BzBzb, \BpBm, \uubar, \ddbar, \ccbar, \ssbar, and \tautau processes in adequate proportions and also beam background. This sample corresponds to an integrated luminosity of $100$~\invfb.
All data and simulated samples are analysed with the Belle~II software~\cite{Kuhr:2018lps}.

\section{Event selection}
\label{sec:selection}

\begin{table}[htb]
  \begin{tabular}{lc}
    \hline
    Decay &  Selection Criteria\\
    \hline
    \multicolumn{2}{c}{$\Bz$ decays} \\
    \hline
    $\Bz \to \Dm \pip$ & \multirow{4}{*}{ $M_{bc} > 5.2\>\gevcc$ and  $-0.2 < \Delta E < 0.2 \>\gev$} \\
    $\Bz \to \Dm \rho^+$   \\
    $\Bz \to \Dstarm \pip$  \\
    $\Bz \to \Dstarm \rho^+$ \\
    \hline
    \multicolumn{2}{c}{$D$ decays}\\
    \hline
    $\Dstarp \to \Dz \pim$ & $ 0.143  < m_{\Dstarp}-m_{\Dz} < 0.147\>\gevcc$ \\
    \hline
    $\Dm \to \Kp\pim\pim$ &  $ |m-m_{PDG}|<0.015\>\gevcc$ \\
    \hline
    $\Dz \to \Km \pip$  &  \multirow{3}{*}{$|m-m_{PDG}|<0.015\>\gevcc$   } \\
    $\Dz \to \Km \pip \piz$ \\
    $\Dz \to \Km \pip \pip \pim$  \\
    \hline
    \multicolumn{2}{c}{$\rho$ decay}\\
    \hline
    $\rho^+ \to \pip \piz$ & $ |m-m_{PDG}|< 0.10\>\gevcc$ \\
  \hline
  \end{tabular}
\caption{List of hadronic final states used to reconstruct $B$-meson decays.}
\label{tab:channelsandcuts}
\end{table}

The \emph{signal side} \Bd meson candidates are reconstructed as detailed in Table~\ref{tab:channelsandcuts}, where \Mbc refers to the beam-energy-constrained mass, \deltaE to the beam-energy difference~\cite{Kou}, $m$ is the invariant mass of a particle and $m_{PDG}$ is its known mass. Charge conjugation is implied throughout this document. The vertex position of the \emph{tag side} $B$, the other $B$ daughter of the \FourS decay, is determined from the tracks not used to reconstruct the signal candidate; the number of such tracks is required to be two or greater.\\

Pions and kaons are identified as follows:
\begin{itemize}
\item The distance from the charged particle track to the interaction point in the $r-\phi$ plane transverse to the beam is required to be less than 0.50~cm, and the track's $z$ position from the interaction point is less than 3.0 ~cm. A selection based on particle identification likelihoods is used to identify  $\pipm$ and  $\Kpm$.
\item Charged pions coming from the $\rho^\pm$ or directly from the $\Bd$ meson have an additional requirement: the momentum in the \FourS center-of-mass frame of the mother particle must be larger than  $ 0.20 \>\gevc$.
\item $\piz$ candidates are formed from pairs of photons, with each photon having an energy greater than 30 MeV, 80 MeV, or 120 MeV if reconstructed in the barrel region, backward end-cap region, or forward end-cap region, respectively. The angle difference between the momenta of the two photons is required to
be less than $0.90$~radians and less than $1$ radian for the angle in the $r-\phi$ plane. The $\piz$ mass is required to be between $121$ and $142$~\mevcc; subsequently, a mass constrained fit is then performed and only candidates surviving the fit pass through the selection.
\end{itemize}

Suppression of continuum background (\epem \to \uubar, \ddbar, \ssbar, \ccbar) is achieved by requiring the following:
\begin{itemize}

\item Reduced Fox-Wolfram moment $R_2 < $ 0.3~\cite{Kou},
\item Cosine of angle between the thrust axis~\cite{Kou} of the signal \Bz and the thrust axis of rest of event $\cos{\theta_{\rm{thrust}}} < 0.8 $.
\end{itemize}

The fitted region, signal region, and sideband region are defined as follows:
\begin{itemize}
\item  Fit region $\Mbc \in [5.20, 5.29]\,\gev/ c^{2}  $ and $ \deltaE \in [-0.20, 0.15]\,\gev$
\item Signal region:  $\Mbc \in [5.27, 5.29]\,\gev /c^{2}  $ and $ \deltaE \in [-0.07, 0.05]\,\gev$
\item Sideband region: $\Mbc \in [5.20, 5.26]\,\gev /c^{2}  $ and $ \deltaE\in [-0.20, -0.10] \cup [0.10, 1.50]\,\gev $ and $ R_2>0.4 $ and $ \cos{\theta_{\rm{thrust}}} < 0.8$
\end{itemize}

The fraction of events with multiple candidates in the signal region used for the final fit to extract \Bd lifetime is 5\%. No best candidate selection is required and all candidates are kept.

The fit to the signal and tag side vertices is performed as follows.
The \emph{signal side} vertex is reconstructed from the hadronic decays listed in Table~\ref{tab:channelsandcuts} using a  kinematic fit of the whole signal decay chain~\cite{Krohn:2019oer}. The \emph{tag side} vertex position is determined from the tracks not used to reconstruct the signal side with an adaptive vertex fitter that accounts for the bias due to secondary decays~\cite{Rave}.

\section{Time difference between the two \B mesons}
\label{sec:deltaTReco}

The proper time difference $\deltaT = t_{\rm sgn} - t_{\rm tag}$ between the two \B meson candidates ({\it signal} and {\it tag}) is approximated from the distance $\Delta z _{boost}$ between the two reconstructed vertices along the collision boost direction $\vec{\beta}$ as follows;
\begin{equation}
\Delta t = \frac {\Delta z _{\rm {boost}}} {\beta \gamma\, c}.
\label{eq:deltat}
\end{equation}

For each event, the uncertainty $\sigma_{\Delta t}$ on the calculated $\Delta t$ is computed. Candidates are required to pass two additional cuts:
\begin{itemize}
	\item $|\Delta t| < 8 \ps$,
	\item $\sigma_{\Delta t} < 3.5 \ps$.
\end{itemize}

\section{Fit strategy}
\label{sec:fit}

In order to estimate the \Bd lifetime from the selected candidates, an unbinned maximum likelihood fit is performed on the overall \deltaT distribution.
The \deltaT function used in the fit is the sum of three components,

\begin{eqnarray}
\mathrm{\textit{P}}_{\rm all}(\Delta t) =
n_{\rm sgn} \>\mathrm{\textit{P}}_{\rm sgn}(\Delta t,\tau_{\Bd})
+n_{\rm \bbbar} \>\mathrm{\textit{P}}_{\rm \bbbar}(\Delta t, \tau_{\textrm{eff}})
+ n_{\rm cont} \>\mathrm{\textit{P}}_{ \rm cont}(\Delta t),
\label{eq:pdfAll}
\end{eqnarray}
where $n_{\rm sgn}, n_{\rm \bbbar} $, and $n_{\rm cont}$ are the yields, respectively, of signal events (\Bd decaying in one of the hadronic decays listed in Table \ref{tab:channelsandcuts} and correctly reconstruced), \bbbar events (\BzBzb or \BpBm wrongly reconstructed as signal decays) and continuum events $(u\bar{u}, c\bar{c}, d\bar{d}, s \bar{s}$ wrongly reconstructed as signal decays).  $\>\mathrm{\textit{P}}_{\rm sgn},  \>\mathrm{\textit{P}}_{\rm \bbbar}, \>\mathrm{\textit{P}}_{\rm cont} $ are consequently the signal, \bbbar, and continuum probability density functions, respectively, used in the unbinned likelihood fit.

An important ingredient of the fit to extract the \Bd lifetime $\tau_{\Bd}$ is the resolution function $\calR$ that describes the smearing impacting the\textit{ \deltaT} measurement due to various experimental effects.
In this analysis $\calR$ is modeled by the sum of three Gaussian distributions:

\begin{eqnarray}
\begin{aligned}
\mathrm{\calR}(\delta_{\Delta t}) =   f_{1} \>{\cal{G}}(\delta_{\Delta t}; \mu_{1},\sigma_{1} )
+f_{2} \>{\cal{G}}(\delta_{\Delta t}; \mu_{1} + \Delta\mu ,s_{2} \times \sigma_{1}) \\
+(1-f_{1} -f_{2})\>{\cal{G}}(\delta_{\Delta t}; \mu_{1},s_{2} \times s_{3}\times \sigma_{1}),
\end{aligned}
\label{eq:3gauss}
\end{eqnarray}

where $\>{\cal{G}}(x;\mu,\sigma)$ denotes a Gaussian distribution with mean $\mu$ and width $\sigma$ and $\delta_{\Delta t}$ is the difference between the reconstructed and the generated $\Delta t$.

\begin{figure}[!tbp]
	\centering
	\includegraphics[width=0.49\textwidth]{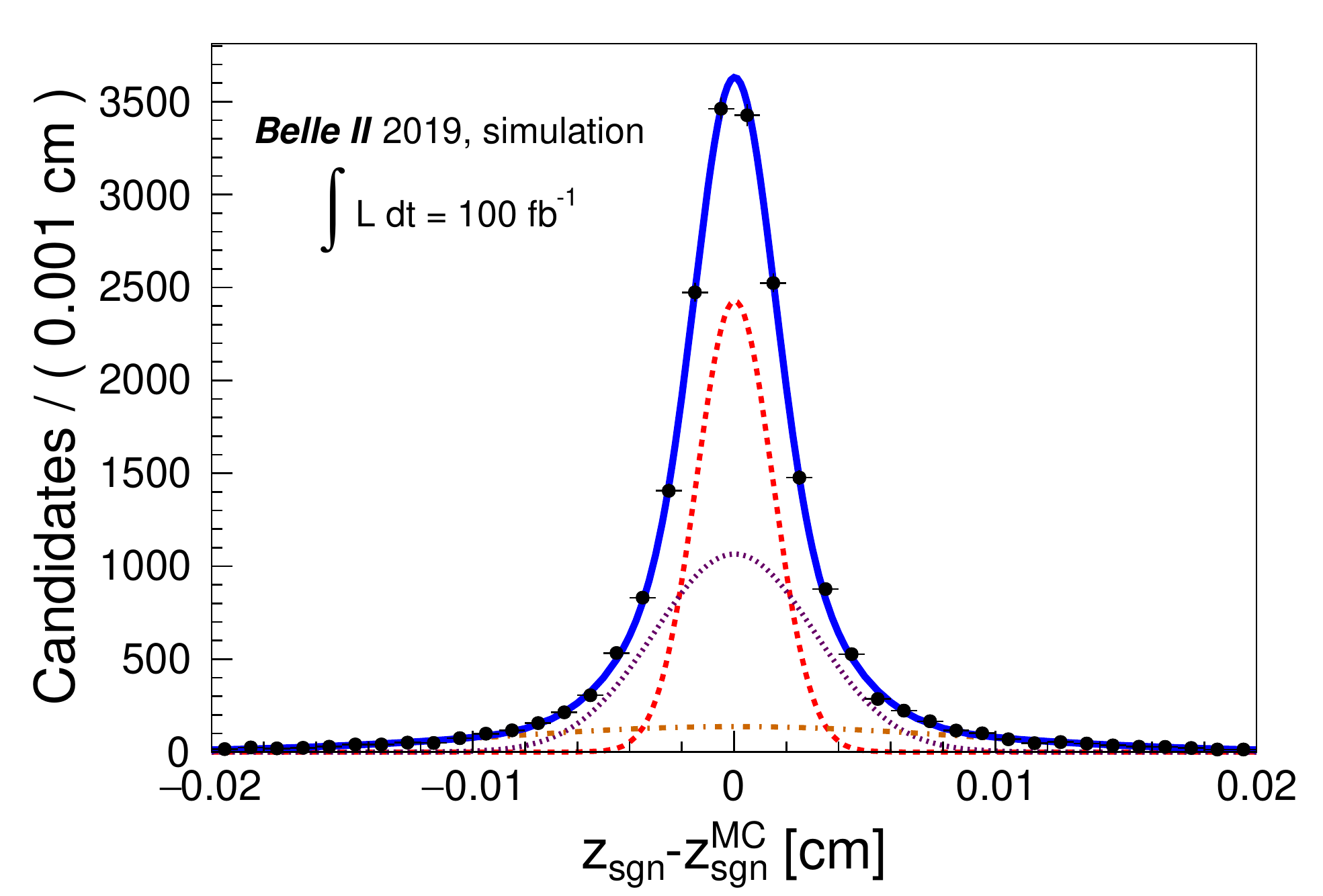}
	\hfill
	\includegraphics[width =0.49\textwidth]{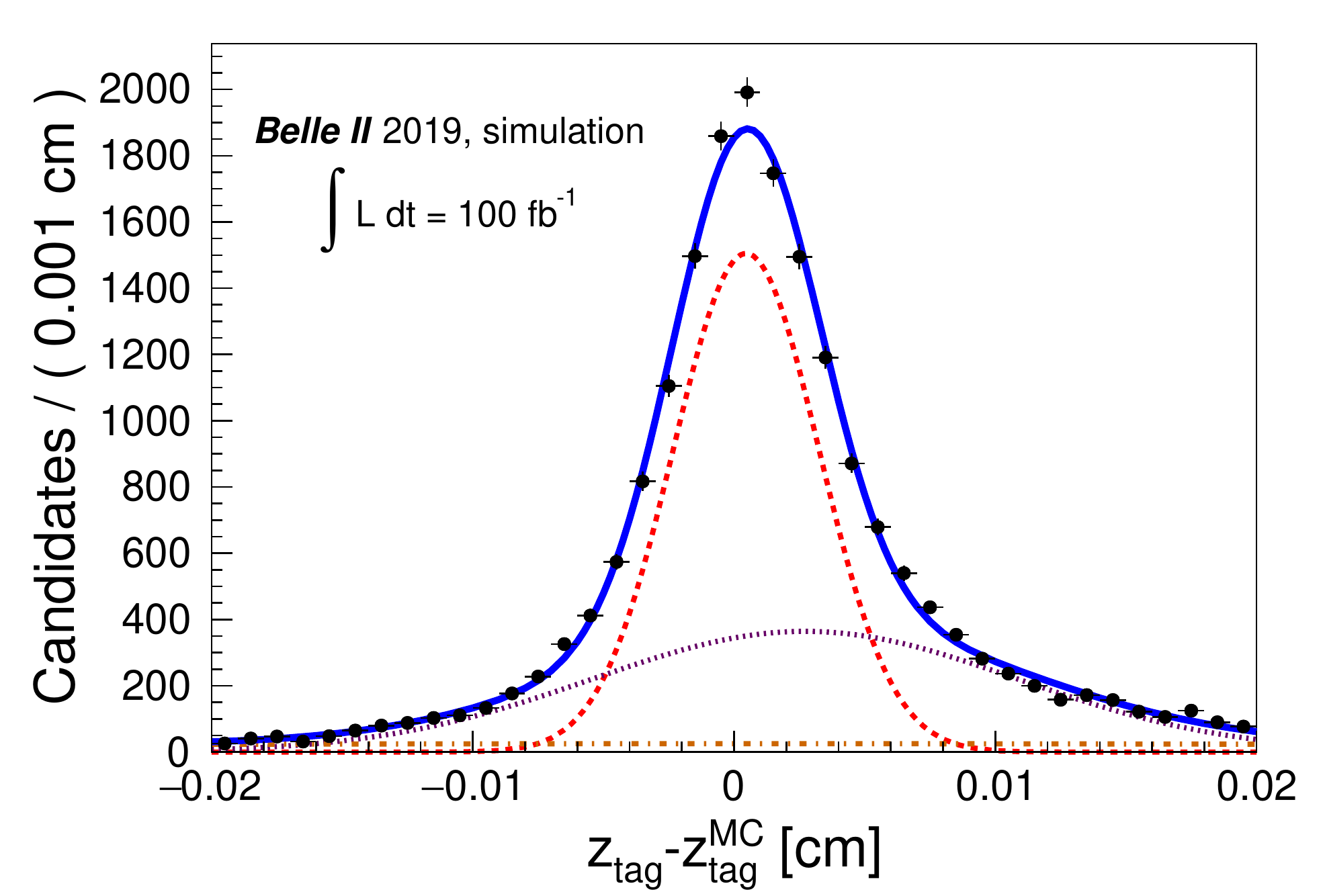}
	\caption{Fit to the $z$-vertex residual for (\textit{left}) the signal $B$ and (\textit{right}) the tag $B$, as obtained from the simulated signal sample. The distributions refer to the true simulated $\Bd \to D^{(*)}h$ decays reconstructed and selected as described in Section~\ref{sec:selection}.}
	\label{fig:resolutions}
\end{figure}

 $\calR$ is also used to describe the $z$-vertex residual, the distribution of the difference between reconstructed and generated signal or tag vertex $z$ position, shown in Figure ~\ref{fig:resolutions}.
The weighted sigma obtained from this latter fits is $\sim 30\,\mu$m for the \emph{signal side} \B and $\sim 80\,\mu$m for the \emph{tag side} \B. These values correspond to an expected \deltaT resolution of $\sim 1$~ps.
The bias toward positive values observed for the residual on the tag side is due to the contribution from secondary decays.

\begin{figure}[tbp]
	\includegraphics[width=0.48\columnwidth]{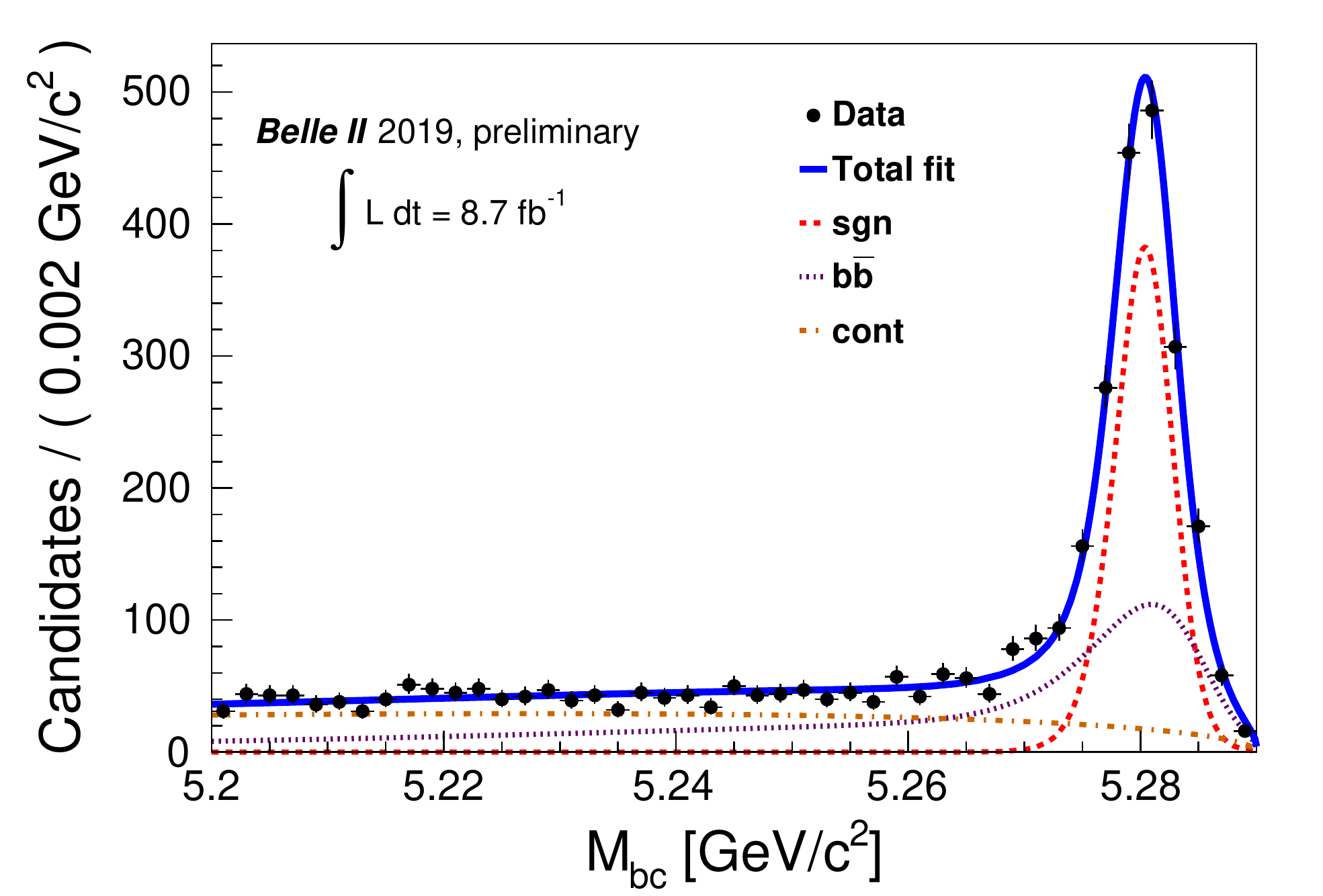}
	\includegraphics[width=0.48\columnwidth]{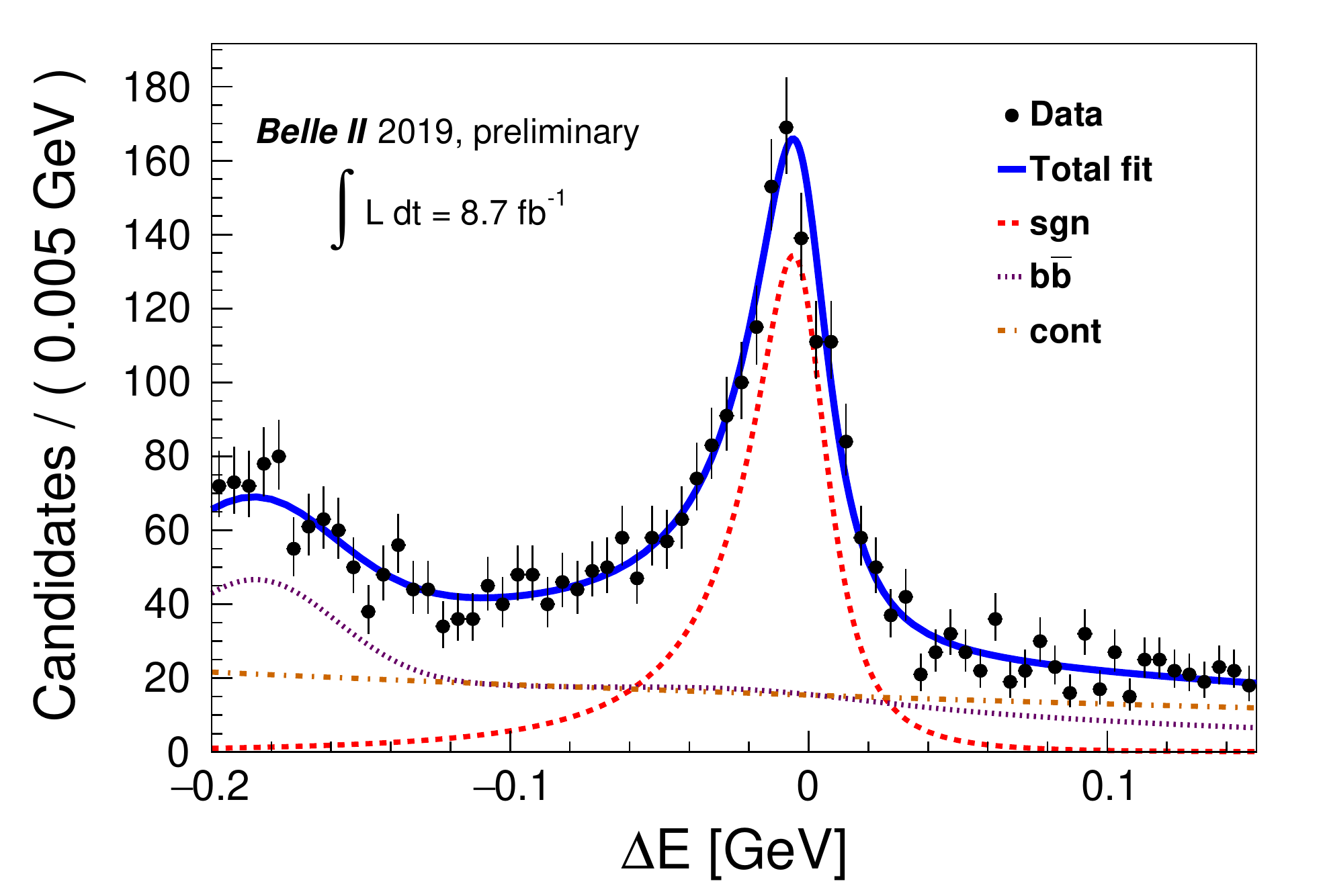}\\
	\caption{Projections of the \textit{\Mbc--\deltaE} fit to the data sample to extract the yields of the different components, see Table \ref{tab:fit_data}: (\textit{left}) \textit{\Mbc}, (\textit{right})\textit{ \deltaE}. The yellow, dark blue, and red lines present the fit contributions corresponding to continuum, \bbbar, and signal components, respectively. The sum of all fit components is presented in light blue. The distributions refer to all $\Bd \to D^{(*)}h$ decay candidates selected as described in Section~\ref{sec:selection}.}
	\label{fig:2Dfit_data}
\end{figure}

\begin{figure}[tbp]
	\includegraphics[width =0.49\textwidth]{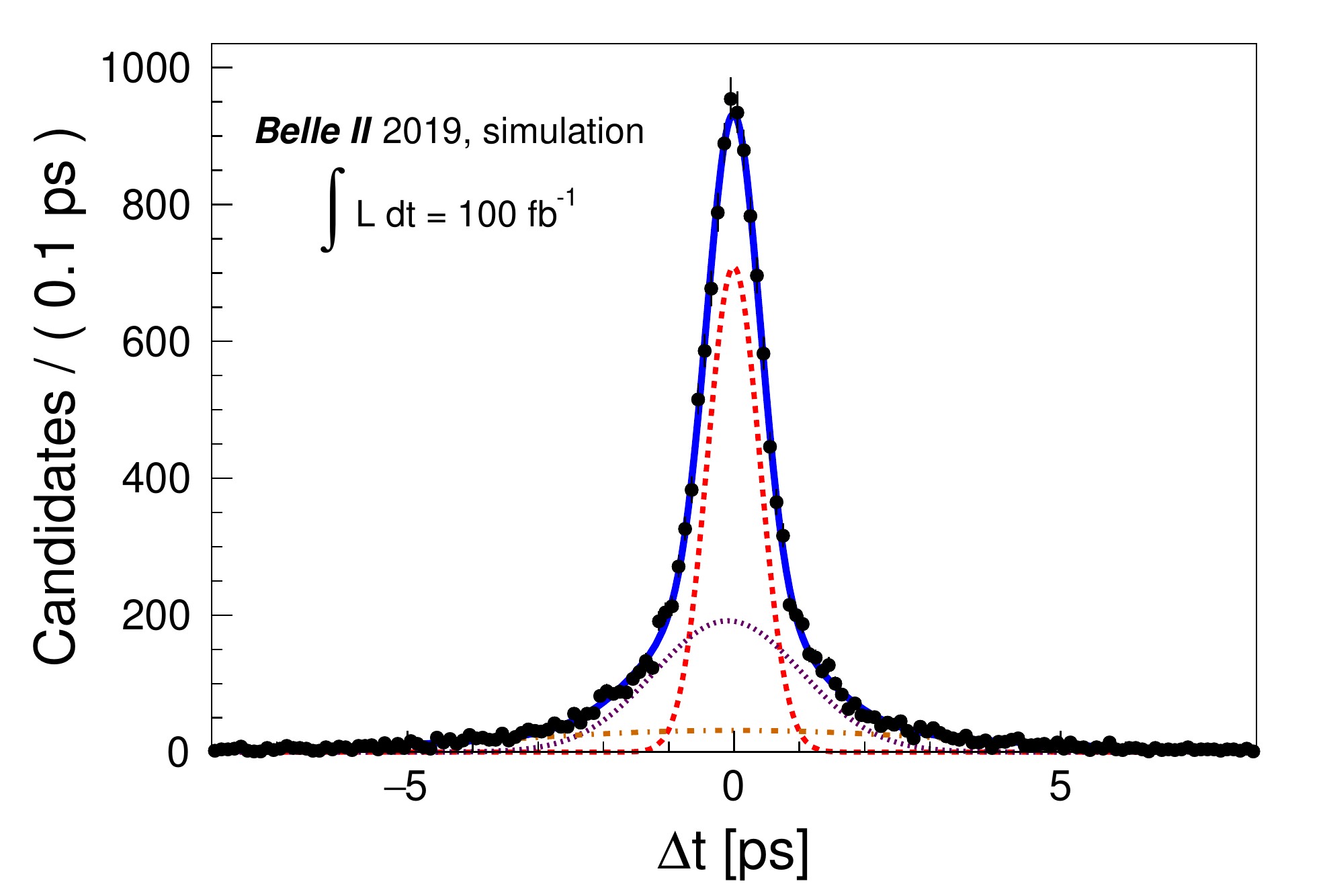}
	\hfill
	\includegraphics[width =0.49\textwidth]{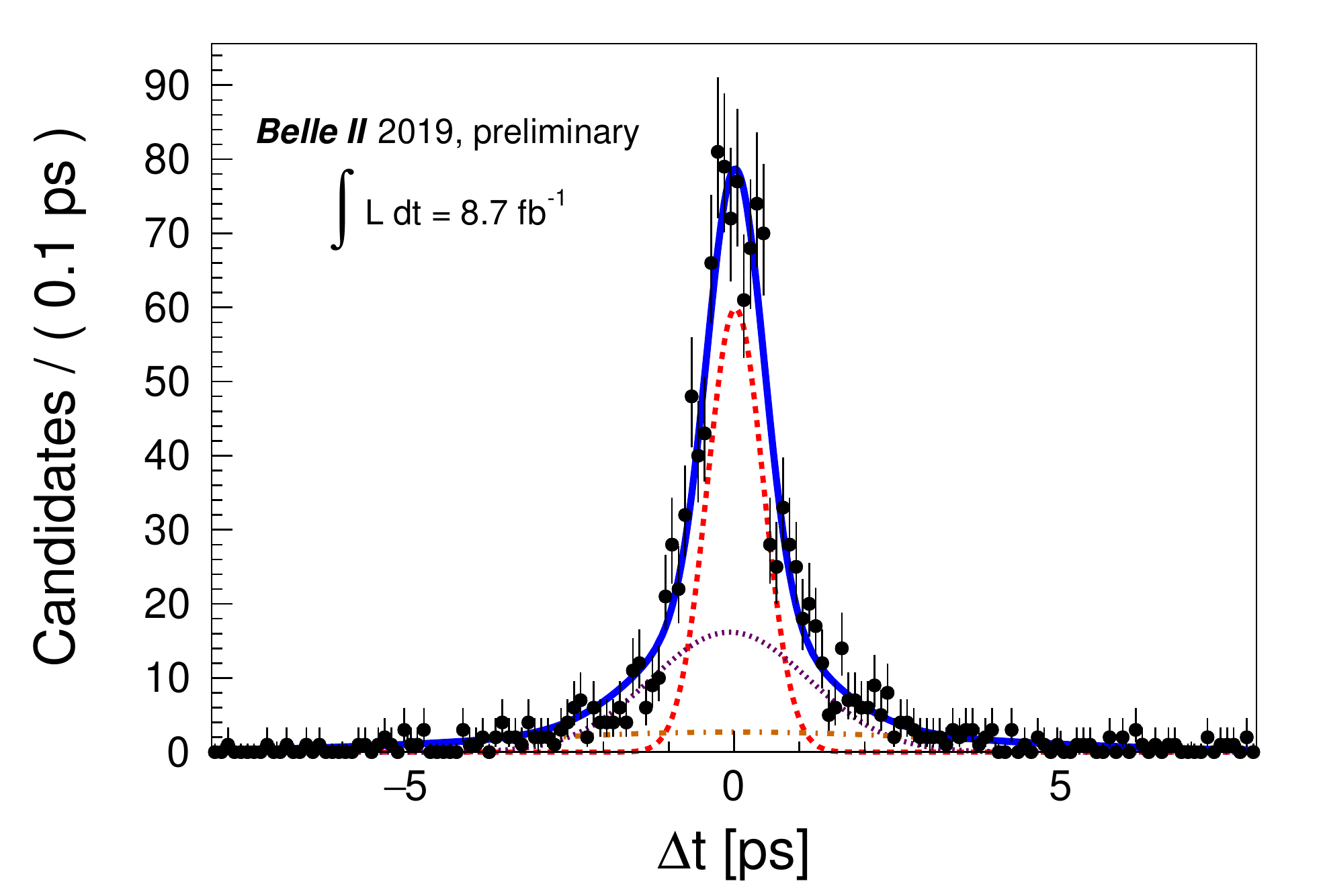}
	\caption{Fit to the $\Delta t$ distribution of the sideband events in (\textit{left}) simulation and (\textit{right}) data. The shape extracted from these fits, with parameters listed in Tables \ref{tab:fixed_params} and \ref{tab:fit_data}, is used to describe the \textit{\deltaT} distribution of the continuum component. The distributions refer to all $\Bd \to D^{(*)}h$ decay candidates selected as described in Section~\ref{sec:selection}.}
	\label{fig:fit_dt_cont}
\end{figure}

\begin{figure}[tbp]
	\centering
	\includegraphics[width=0.7\textwidth]{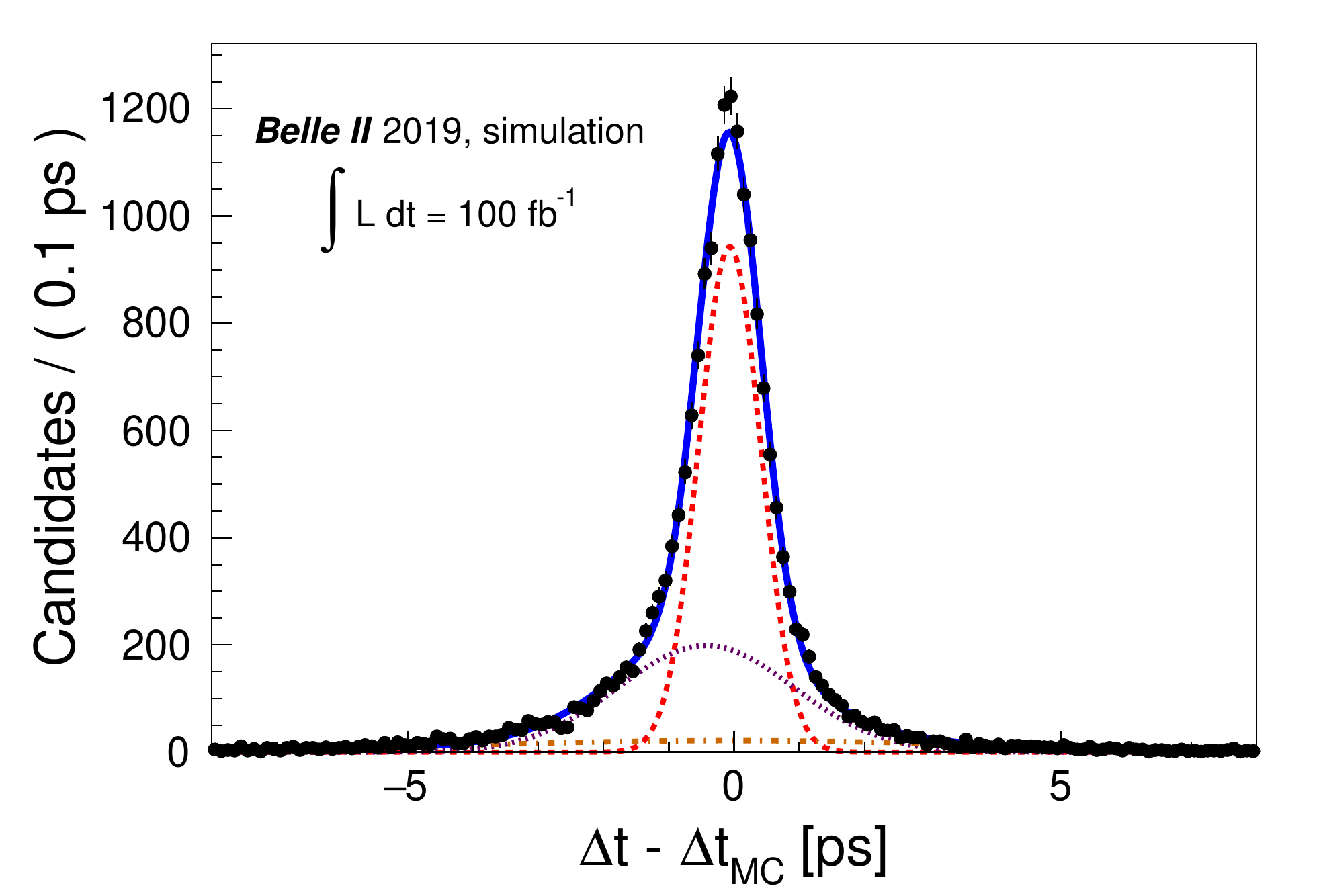}
	\caption{Fit to the \deltaT residual for the signal contribution of the simulated sample leaving all the parameters free.
		These parameters listed in Table \ref{tab:fixed_params}, except $\mu_{1_{res}} $and $ \sigma_{1_{res}} $, are fixed for \calR in the \Bd lifetime fit. The distributions refer to the true simulated $\Bd \to D^{(*)}h$ decays reconstructed and selected as described in Section~\ref{sec:selection}.}
	\label{fig:fit_res_sgn}
\end{figure}

\begin{figure}[tbp]
  \includegraphics[width=0.85\columnwidth]{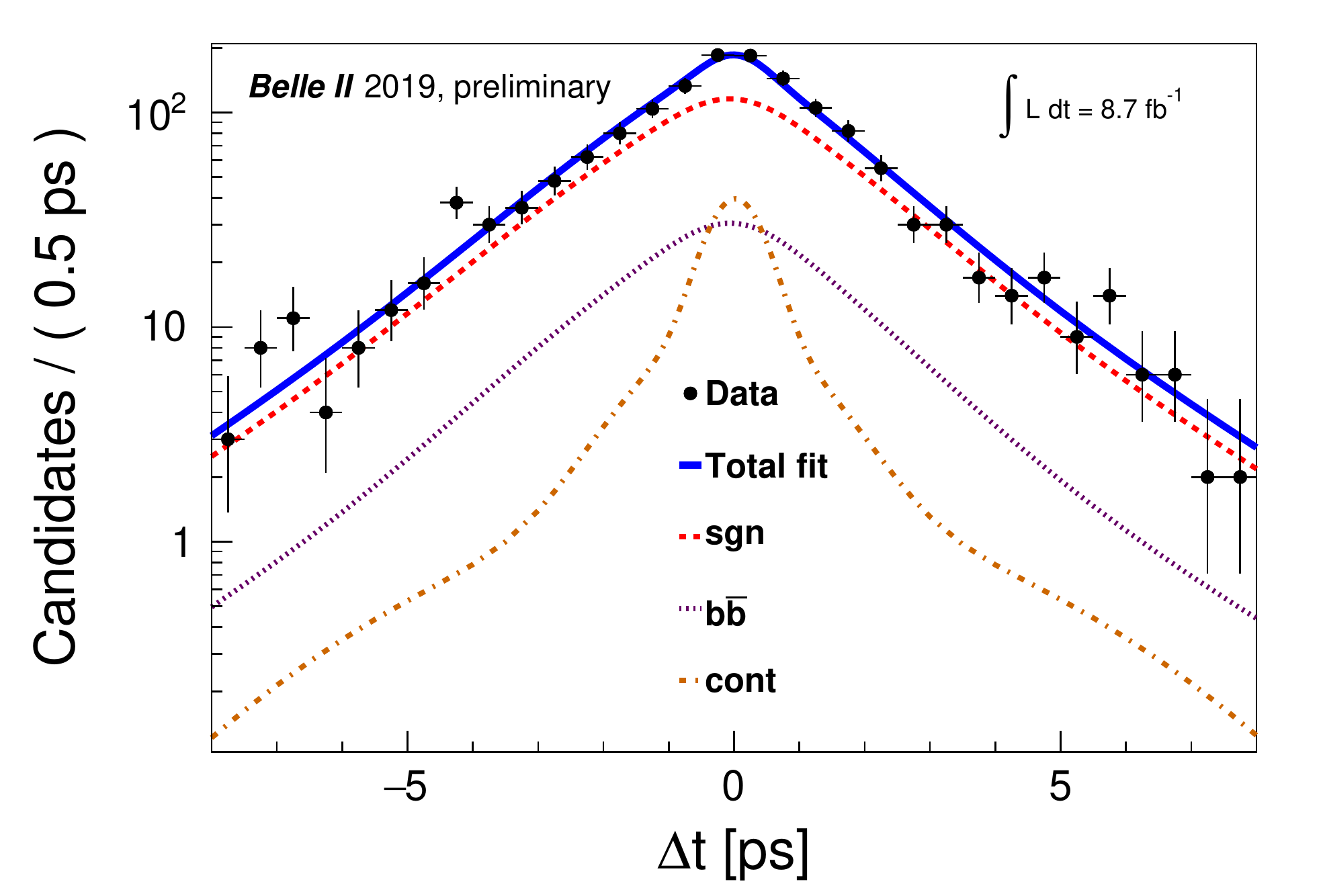}
  \includegraphics[width=0.85\columnwidth]{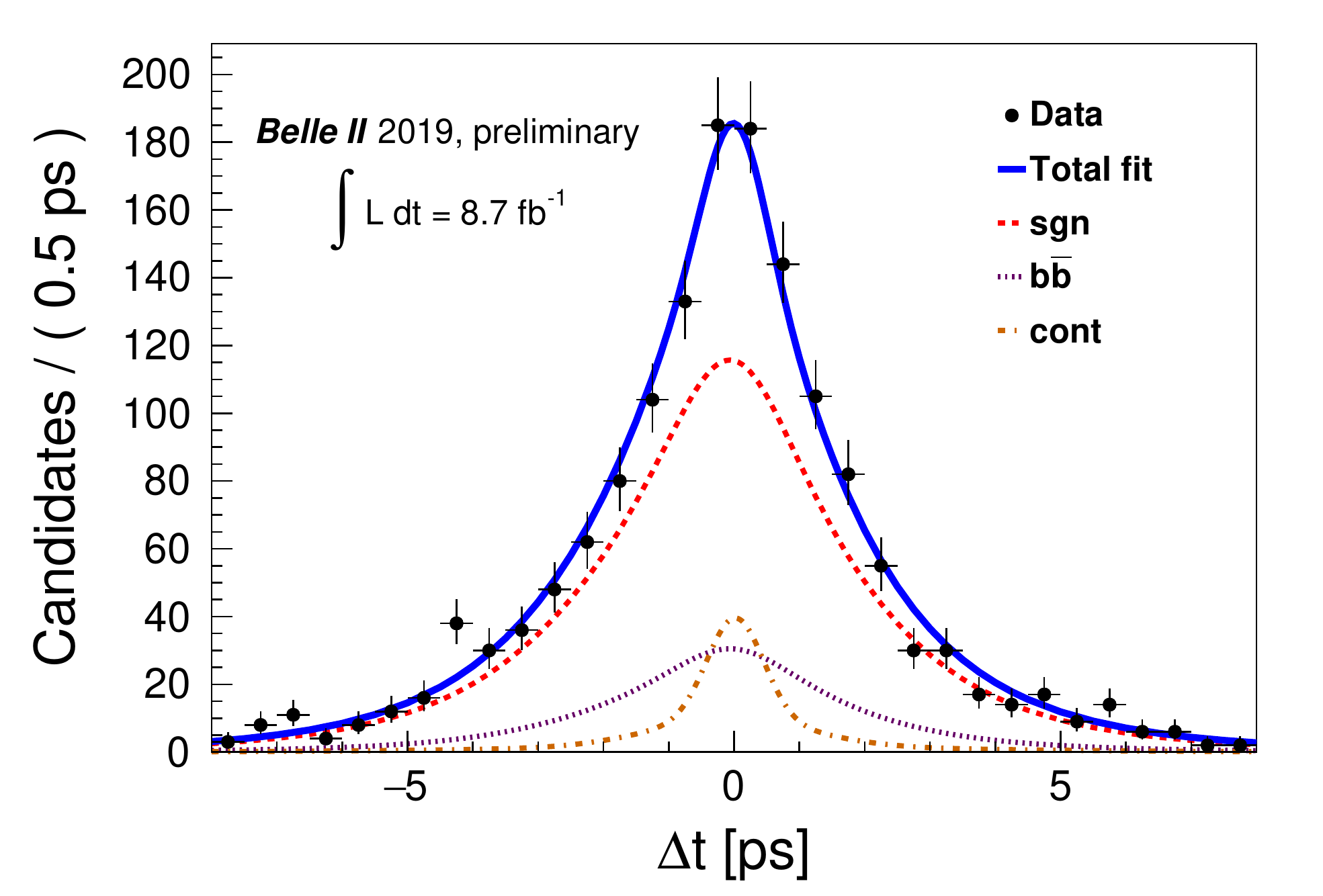}
  \caption{Fit to $\Delta t$ distribution for all $\Bd \to D^{(*)}h$ selected decay candidates in the 2019 Belle~II data set (logarithmic scale on top and linear scale on bottom). The yellow, dark blue, and red lines present the fit contributions corresponding to continuum, \bbbar and signal components, respectively. The sum of all fit components is presented in light blue. See Sections~\ref{sec:selection} and \ref{sec:fit} for explanations on the selection and fit procedures, respectively.}
  \label{fig:fullfit_data}
\end{figure}

The different contributions to Equation~(\ref{eq:pdfAll}) are obtained from three separate fits as follows.
\begin{itemize}
\item[$\bullet$ {\bf Fit 1}] $n_{\rm sgn}, n_{\rm \bbbar} $, and $n_{\rm cont}$ are obtained from a two-dimensional extended maximum likelihood fit to the unbinned \textit{\Mbc--\deltaE} distributions (Figure~\ref{fig:2Dfit_data}).
The yields in the signal region are obtained with an integral of the probability density functions in the signal region.
\item[$\bullet$ {\bf Fit 2}] $ \mathrm{\textit{P}}_{\rm cont}(\deltaT)$ is the sum of three Gaussian distributions having the same functional form as \calR; this assumes no lifetime for the continuum component. In the final fit to extract the \Bd lifetime, its shape is fixed using the parameters obtained from the fits to the data side-band \textit{\deltaT} distributions shown in Figure~\ref{fig:fit_dt_cont}, from simulated and data samples.
\item[$\bullet$ {\bf Fit 3}] $\mathrm{\textit{P}}_{\rm \bbbar}(\deltaT, \tau_{\textrm{eff}})$ and  $\mathrm{\textit{P}}_{\rm sgn}(\deltaT, \tau_{\Bd}) $ are both modeled as an exponential function convolved with the resolution model \calR described in equation \ref{eq:3gauss}. The \bbbar effective lifetime ($\tau_{\textrm{eff}}$) is fixed from a separate fit to only the simulated \bbbar contribution. \calR is shared between $\mathrm{\textit{P}}_{\rm \bbbar}(\deltaT, \tau_{\textrm{eff}})$ and  $\mathrm{\textit{P}}_{\rm sgn}(\deltaT, \tau_{\Bd}) $ in the \Bd lifetime fit. Its parameters, except $\mu_{1res} $ and $ \sigma_{1res}$, are fixed from a fit to the \deltaT signal residuals obtained with the simulated sample and shown in Figure~\ref{fig:fit_res_sgn}.
\end{itemize}
The final fit using Equation~(\ref{eq:pdfAll}) to extract the \Bd lifetime from the \deltaT distribution of all the selected candidates is shown in Figure~\ref{fig:fullfit_data}. In this fit, only three parameters are left floating: $\mu_{1res}, \sigma_{1res}$ and $\tau_{\Bd}$. The obtained lifetime is $\tau_{\Bd}$  = $1.48 \pm 0.28$ ps and the corresponding reduced $\chi^{2} $ for this fit is 0.83. Table~\ref{tab:fixed_params} enumerates the values of the parameters obtained from fits to the simulated sample and that are fixed in the final fit extract the \Bd lifetime on data. All parameters extracted from fits to the data are summarized in Table~\ref{tab:fit_data}. To avoid experimenters’ bias, the fit to the \deltaT distribution on data was performed after the analysis procedure was finalized using simulated samples.

\begin{table}[htb]
  \begin{center}
    \begin{tabular}{l|lc}
      \hline
      & Fit parameter & value \\
      \hline
      \multirow{5}{*}{Fit 2 } & $\Delta\mu_{cont} $ [\gevcc] & $-0.0846 \pm 0.026$ \\
      & $f_{1cont} $  & $0.468  \pm 0.021$ \\
      & $f_{2cont} $ & $0.357 \pm 0.017$ \\
      & $s_{2cont}$ & $ 2.88 \pm 0.096$ \\
      & $s_{3cont}$ & $ 3.02 \pm 0.11 $ \\
      \hline
      \multirow{7}{*}{Fit 3 } &  $\Delta\mu_{res} $ [\gevcc] & $ -0.360 \pm 0.031 $ \\
      & $f_{1res} $ & $ 0.561 \pm 0.016$  \\
      & $f_{2res} $ & $ 0.336 \pm 0.014$ \\
      & $s_{2res}$ & $ 2.84 \pm 0.079$ \\
      & $s_{3res}$ & $ 2.99 \pm 0.13$ \\
      & \multicolumn{2}{c}{\rule{0.4\linewidth}{0.4pt}}\\
      & $\tau_{\textrm{eff}}$ [ps]& $ 1.31 \pm 0.03 $ \\
      \hline
    \end{tabular}
  \end{center}
  \caption{Values of the parameters obtained from unbinned maximum likelihood fits to the simulated sample: Fit 2 on \deltaT distribution for the side-band and Fit 3 on the \deltaT distribution for the signal contribution.}
  \label{tab:fixed_params}
\end{table}

\begin{table}[htb]
  \begin{center}
    \begin{tabular}{l|lc}
      \hline
      & Fit parameter & value \\
      \hline
      \multirow{7}{*}{Fit 1} & $n_{sgn}$ fit region & $\left(1.22 \pm 0.04\right) \times 10^{3}$\\
      & $n_{b\overline{b}}$ fit region & $\left(1.29 \pm 0.07\right) \times 10^{3}$\\
      & $n_{cont}$ fit region & $\left(1.14 \pm 0.06\right) \times 10^{3}$\\
      & \multicolumn{2}{c}{\rule{0.3\linewidth}{0.4pt}}\\
      & $n_{sgn}$ signal region & $\left(1.10 \pm 0.04\right) \times 10^{3}$\\
      & $n_{b\overline{b}}$ signal region & $270 \pm 32$\\
      & $n_{cont}$ signal region & $140 \pm 21$\\
      \hline
      \multirow{2}{*}{Fit 2} & $\mu_{1cont}$  [ps] & $0.021 \pm 0.021$\\
      & $\sigma_{1cont}$  [ps] & $0.429 \pm 0.015$\\
      \hline
      \multirow{3}{*}{Final fit }  & $\mu_{1res}$ [ps] & $-0.03 \pm 0.06$\\
      & $\sigma_{1res}$  [ps] & $0.56 \pm 0.18$\\
      & $\tau_{\Bz}$  [ps] & $1.48 \pm 0.28$\\
      \hline
    \end{tabular}
  \end{center}
  \caption{Parameters extracted from unbinned maximum likelihood fits to the data: Fit 1 on \Mbc--\deltaE 2D distribution, Fit 2 on \deltaT distribution for the side-band and the Final fit on the \deltaT distribution in the signal region. }
  \label{tab:fit_data}
\end{table}

\section{Systematic studies}
\label{sec:systematics}

The limited size of the early data set on which this measurement is based, from Section~\ref{sec:dataset}, yields a statistical uncertainty of about 20\%, while the \Bd lifetime is known with a 0.3\% relative uncertainty. Consequently, systematic studies focused on the two main expected uncertainty sources: the
parametrizations used in the fits described in Section~\ref{sec:fit} (partly fixed from simulated data) and the vertices reconstruction (embedding potential alignment and calibration effects).

The first study relies on the reproduction of the full fit strategy for data described in Section~\ref{sec:fit} over ten thousand simulated data samples with sizes equivalent to the recorded data sample. Those samples were obtained with the bootstrapping technique from a fully simulated sample ten times larger.
From the distribution of the ten thousand fitted lifetimes, a systematic uncertainty of $0.05$~ps is obtained. Moreover, the final fit is repeated leaving also $\tau_{\textrm{eff}}$ free, resulting in a further systematic uncertainty of $0.01$~ps due to the difference with respect to the nominal fit of the \Bd lifetime.

In a second step, the complete analysis is reproduced on a subset of the real data, which was processed with
two different calibration and alignment constants to mimic the effect of the residual misalignment and miscalibration. A systematic uncertainty of $0.03$~ps is extracted from the difference between the lifetime values obtained.

The overall systematic uncertainty is obtained as the quadratic sum of these contributions and amount to $0.06$~ps or 4\% relative uncertainty.

\section{Conclusion}

A first measurement of the \Bd lifetime was performed with the 2019  Belle~II data set, which amounts to an integrated luminosity of $8.7\pm0.2$~\invfb. The estimated lifetime is \hbox{$\tau_{\Bd} = 1.48 \pm 0.28 \pm 0.06$} ps, where the first uncertainty is statistical and the second is systematic. This value is compatible with the world average of $1.519 \pm 0.004$~ps~\cite{pdg}.

\FloatBarrier

\bibliography{bibliography}

\end{document}